\documentclass[12pt]{article}
\usepackage[utf8]{inputenc}
\usepackage[margin=3cm]{geometry}
\usepackage{amsmath, amsfonts, amssymb}
\usepackage{color, hyperref}
\usepackage{graphicx, epstopdf}

\newcommand{\be}{\begin{equation}}
\newcommand{\ee}{\end{equation}}
\newcommand{\bea}{\begin{eqnarray}}
\newcommand{\eea}{\end{eqnarray}}

\begin{document}

\begin{titlepage}

\begin{flushright}
{\small
DESY 21-224}
\end{flushright}
\vspace{.3in}

\begin{center}
{\Large\bf
A sonic boom in bubble wall friction\\}%
\vspace{1cm}{
  {\large
Glauber C.~Dorsch$^a$,
Stephan J.~Huber$^b$,
Thomas Konstandin$^c$
}}

{\small\vskip5mm
$^a$ Universidade Federal de Minas Gerais,  31270-901, Belo Horizonte, MG, Brazil \\
$^b$ University of Sussex, Brighton, BN1 9QH, UK \\
$^c$ DESY, Notkestra{\ss}e 85, 22607 Hamburg, Germany\\
}

\bigskip

\begin{abstract}
We revisit the computation of bubble wall friction during a cosmological first-order phase transition, using an extended fluid \emph{Ansatz} to solve the linearized Boltzmann equation. A singularity is found in the fluctuations of background species as the wall approaches the speed of sound. Using hydrodynamics, we argue that a discontinuity across the speed of sound is expected on general grounds, which manifests itself as the singularity in the solution of the linearized system. We discuss this result in comparison with alternative approaches proposed recently, which find a regular behaviour of the friction for all velocities.
\end{abstract}

\end{center}

\end{titlepage}

\section{Introduction}

The recent detection of gravitational waves, and the forthcoming launch of the space-based interferometer LISA, have recently boosted the interest on the dynamics of phase transitions in the early Universe and their observational consequences.
If the phase transition is first order, it will proceed via bubble nucleation, expansion and percolation, producing sound waves and turbulence in the plasma which, together with the kinetic energy of the Higgs field, could source stochastic gravitational waves observable by LISA~\cite{Caprini:2015zlo, Caprini:2019egz}. Other relics could also be produced during this process, such as a dark matter abundance~\cite{Azatov:2021ifm} and a matter-antimatter asymmetry~\cite{Kuzmin:1985mm,Morrissey:2012db, Konstandin:2013caa}, and hence the detection of primordial radiation by LISA could also give us information on the underlying particle physics model. Crucially, the abundance of relics produced by this phase transition depends, among other few parameters, on the expansion velocity of these bubbles. The determination of this velocity depends on the microphysics of the interactions of the bubble wall with particles in the plasma. The passage of the bubble will drive the plasma out of equilibrium, and this will backreact on the bubble, acting as a dissipative friction force which tends to halt its expansion. Thus, an accurate determination of these relics, especially the stochastic gravitational wave spectrum generated at the early Universe, relies on an adequate modelling of this out-of-equilibrium dynamics between the Higgs field and the hot plasma.

The calculation of this plasma friction has been previously tackled in the literature using the so-called \emph{fluid Ansatz} for the non-equilibrium particle distribution function~\cite{Moore:1995si, Konstandin:2014zta}. This approach describes non-equilibrium via three fluctuations --- chemical potential, temperature and velocity fluctuations ---, each with a specific momentum dependence. This allows for the computation of the collision terms (at least in a leading-log approximation), thus converting the integro-differential Boltzmann equation into a more manageable set of ODE's for the three parameters appearing in the \emph{Ansatz}. It is then found that the friction becomes singular as the wall velocity approaches the speed of sound in the plasma~\cite{Moore:1995si, Konstandin:2014zta} --- an effect akin to a sonic boom. Furthermore, the formalism also predicted that baryogenesis would be utterly impossible at any supersonic wall velocity. 

In fact, both effects have one and the same origin: a vanishing eigenvalue in the kinetic matrix of the Boltzmann system when the wall velocity equals the sound speed. At this singular point, all damping rates become increasingly large, meaning that all modes which tend to be brought back to equilibrium by mutual collisions will do so very quickly. Baryogenesis then vanishes because it cannot work in equilibrium. However, energy-momentum conservation requires some of the fluctuations of the light species (such as photons, gluons and light fermions) to actually be undamped, and in this case the vanishing eigenvalue in the kinetic matrix will translate to a discontinuity in the temperature and velocity of the plasma across the wall boundary. The need for such a discontinuity is in fact confirmed by macroscopic hydrodynamical analyses~\cite{1994A, 1994B, 1995C,Espinosa:2010hh}.

Recently, however, an alternative approach for treating the dynamics of microscopic non-equilibrium has been put forward~\cite{Cline:2020jre}, claiming to solve the Boltzmann equation without specifying any \emph{Ansatz} for the particle abundances~\footnote{For other recent developments on bubble wall friction see~\cite{BarrosoMancha:2020fay,Hoche:2020ysm,Azatov:2020ufh,Balaji:2020yrx,Cai:2020djd,Wang:2020zlf, Bigazzi:2021ucw, Ai:2021kak, Baker:2021sno, Lewicki:2021pgr, Gouttenoire:2021kjv}.}. Instead, the authors introduce an implicit truncation scheme in the mode expansion of the system, assuming that all higher moments of the distribution function are proportional to the first order fluctuation in velocities. It is then shown that the formalism predicts a continuous behaviour for observables at all wall velocities, including across the speed of sound boundary. In particular it predicts that electroweak baryogenesis can even work with supersonic walls, even if at a somewhat reduced rate. When applied to a study of the friction and the wall velocity, the formalism again predicted a continuous behaviour~\cite{Laurent:2020gpg}\footnote{Recent applications of this approach can be found in~\cite{Cline:2021iff, Lewicki:2021pgr}.}. 

One is then led to wonder about the origin of such discrepancies and the reliability of two different methods that yield so contrasting results. For the baryogenesis computation this issue has been recently tackled in~\cite{Dorsch:2021ubz}, where it was shown that supersonic baryogenesis is indeed possible (even if suppressed) if one considers an \emph{extended fluid Ansatz} with more than just three perturbations. Moreover, a continuous behaviour does indeed emerge across all velocities as more and more perturbations are included in the description, thus bringing the predictions of the fluid \emph{Ansatz} in agreement with those of the \emph{Ansatz}-less approach of~\cite{Cline:2020jre}.

However, it still remains to be settled whether the two approaches would lead to similar predictions for the friction calculation as well, i.e. whether the discontinuity predicted in the three-fluid approximation for the background fluctuations could be brought under control by extending the \emph{Ansatz} to include more perturbations. The goal of the present paper is to assess this question. It will be found that, unfortunately, the answer is negative.

Even though the friction in the bubble wall during the phase transition constitutes a quite similar problem to baryogenesis, there are some notable differences. One main difference being that, for the friction calculation, one is interested in CP-even deviations from equilibrium, whereas for baryogenesis they are CP-odd, and this leads to some qualitative differences in the collision terms. Furthermore, energy-momentum conservation plays a prominent role in the friction calculation. At the same time, as mentioned above, the singularity in the friction close to the speed of sound is caused by essentially the same reason that suppresses the baryon asymmetry for supersonic walls: some of the eigenvalues of the linearized fluid system change sign as the wall speed becomes supersonic. For baryogenesis, this leads to the fact that deviations are only non-zero behind that wall, thus suppressing the BAU. For the friction calculation the sign change of the eigenvalue results in a divergence of the friction (and hence the breakdown of the perturbative scheme). 

In the present work, we will argue that the discontinuity in the friction calculation is actually physical. We will study 
in detail the origin of the singularity of the lineralized system. We give some general arguments why such a `sonic boom' should occur 
in the plasma for wall velocities close to the speed of sound. Finally, we support our arguments with an
explicit calculation using higher moments and a generalized Ansatz in the Boltzmann equation, as done in~\cite{Dorsch:2021ubz} for the 
analysis of the baryon asymmetry.

The paper is organized as follows. In section~\ref{sec:origin} we review the hydrodynamics of the phase transition to show that a discontinuity in temperature and fluid velocity across the bubble wall is expected on the grounds of energy-momentum conservation. Our point is that the macroscopic analysis confirms the discontinuity predicted by the microphysical approach with the fluid \emph{Ansatz}. In section~\ref{sec:setup} we review the friction calculation with the three-fluid \emph{Ansatz} of~\cite{Konstandin:2014zta} and discuss how the discontinuity emerges in this approach. The extended fluid \emph{Ansatz} is then reviewed in section~\ref{sec:gen_fluid}, and the friction calculation is performed in section~\ref{sec:friction}, where we also show our main results. Finally, section~\ref{sec:conclusions} is left for discussions and conclusions.

\section{The origin of the discontinuity}
\label{sec:origin}

Before we study the friction calculation using an extended Ansatz for the 
fluid, we will provide some general arguments based on hydrodynamic considerations. 

As we will see later, in the linearized system 
the origin of the singularity/discontinuity in the friction arises from a sign change in the eigenvalues of the Liouville operator in 
conjunction with energy-momentum-conservation. This system can also be studied on length scales much larger than
the longest damping scale in the problem. Since the wall thickness is many orders smaller than the 
bubble size, this separation of scales is easily achieved. This system is then in local equilibrium 
and fulfills the hydrodynamic equations. 

This hydrodynamic setup is typically studied to 
analyze the overall energy budget of the phase transition and the expansions mode. In the following we follow the conventions of~\cite{Espinosa:2010hh}.

Consider a system where the plasma in both phases is described by a radiation component  and a bag constant $\epsilon$ in one of the two phases. The pressure of the system then reads 
\be
p_\pm = \frac{a_\pm}{3} T^4 - \epsilon_\pm  \, .
\ee
Defining the enthalpy $\omega = e + p$, conservation of the energy momentum tensor 
\be
T_{\mu\nu} = u_\mu u_\nu \omega - g_{\mu\nu} p
\ee
then reads (in the wall frame) 
\be
\Delta T_{0z} = \Delta T_{zz} = 0,
\label{eq:Tconserv}
\ee
or explicitly 
\be
\gamma_+^2 v_+^2 \omega_+ + p_+ = 
\gamma_-^2 v_-^2 \omega_- + p_- \, ,\quad 
\gamma_+^2 v_+ \omega_+ = 
\gamma_-^2 v_- \omega_- \, .
\ee
A change in the bag constant will be accompanied with 
a change in the plasma temperature and velocity, $T_+ \not= T_-$ and $v_+ \not= v_-$. 
Using the notation $\alpha = (\epsilon_+ - \epsilon_-) / (a_+ T_+^4)$   and eliminating the temperatures then leads to the relation~\cite{Espinosa:2010hh} 
\be
v_+ = \frac{1}{1+\alpha}
\left[
\left( \frac{v_-}{2} + \frac{1}{6v_-} \right) \pm \sqrt{
\left( \frac{v_-}{2} + \frac{1}{6v_-} \right)^2  + \alpha^2 + \frac23 \alpha - \frac13
}
\right] .
\label{eq:vpvm}
\ee
The equation has per se two solutions but only one solution is consistent globally. For wall velocities beyond the Jouguet velocity, the fluid in front of the wall has to be at rest in the plasma frame. This implies $v_+ > v_-$ and is called a detonation. For wall velocities below the 
speed of sound, the fluid behind the wall has to be at rest in the plasma frame. This implies $v_+ < v_-$ and is called a deflagration. For wall velocities in between, hybrid solutions are the most likely outcome which technically are also deflagrations, $v_+ < v_-$. See~\cite{Espinosa:2010hh} for details.

The relation (\ref{eq:vpvm}) has some interesting properties. One can rewrite it as 
\be
v_+ = \frac{1}{1+\alpha}
\left[
X_+ \pm \sqrt{
X_-^2  + \alpha^2 + \frac23 \alpha 
}
\right] 
\label{eq:vpvmrewrite}
\ee
using $X_\pm = v_- /2 \pm 1/(6v_-)$. For $\alpha=0$, the relation reads $v_+ = X_+ \pm X_-$, which yields $v_+ = v_-$ and $v_+ v_- = 1/3$.
For $v_- \to c_s = 1/\sqrt{3}$, one obtains $X_+ \to c_s$ and $X_- \to 0$, which translates into 
\be
v_+ = \frac{1}{1+\alpha}
\left[
c_s \pm \sqrt{
\alpha^2 + \frac23 \alpha 
}
\right] \quad {\rm for} \quad (v_- \to c_s).
\label{eq:cslimit}
\ee
However, in the limit of very weak phase transitions and expanding in $\alpha$ one obtains
\be
v_+ \simeq \frac{1}{1+\alpha}
\left[
X_+ \pm X_- 
\pm \frac13 \frac{\alpha}{X_-} 
\right] \quad {\rm for} \quad (\alpha \to 0).
\label{eq:vpvmsing0}
\ee
Selecting the physical branch then leads in leading order
\be
v_- - v_+ =   \frac{3 v_- (1 - v_-^2) }{1 - 3 v_-^2} \, \alpha \, ,
\quad {\rm for} \quad (\alpha \to 0).
\label{eq:vpvmsing}
\ee
\begin{figure}
    \centering
    \includegraphics[width=0.5\textwidth]{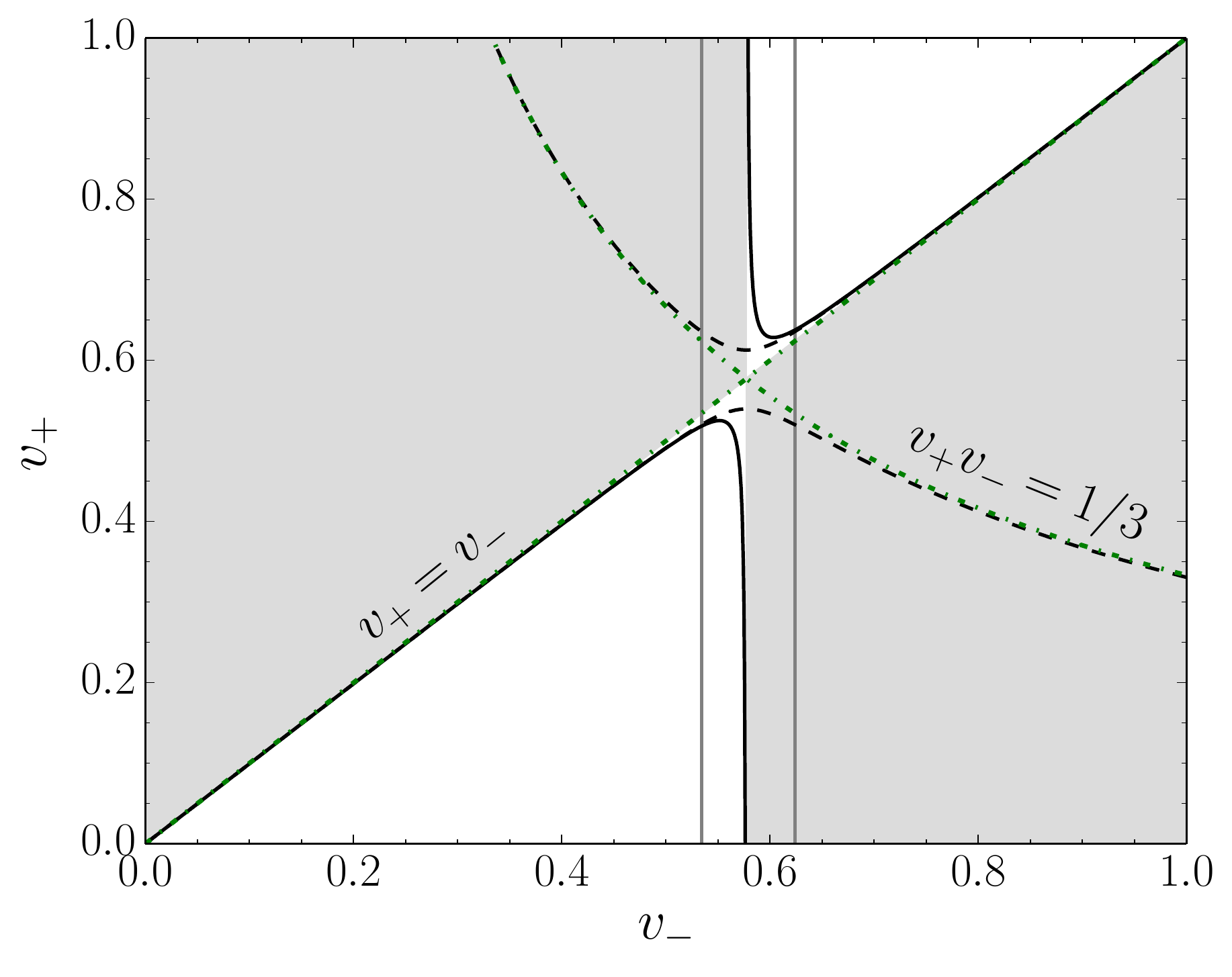}
    \caption{The exact (black dashed) and linearized (black solid) curves for the relation between $v_+$ and $v_-$. The gray shaded region are unphysical. For the linearized relation only the physical branch is shown, and the discontinuity across the speed of sound is clear. The strength of the phase transition is $\alpha = 0.002$. The region in between the vertical lines in the middle satisfies $\alpha/X_-^2>1$, where we expect a breakdown of the linearization procedure. Indeed, it is clear that, outside  this region, the solid and dashed curves agree.}
    \label{fig:vmvp}
\end{figure}%
This is basically the expression one will necessarily obtain once the Boltzmann equations are linearized and 
expanded around the equilibrium. Unfortunately, this limit shows a divergence for $X_- \to 0$ or $v_- \to c_s$.
The divergence stems from the fact the the square root in (\ref{eq:vpvm}) is non-analytic in this limit.
Figure \ref{fig:vmvp} shows the exact and linearized relations between $v_+$ and $v_-$, see (\ref{eq:vpvm}) and (\ref{eq:vpvmsing}). The strength of the phase transition is $\alpha = 0.002$ (cf. section~\ref{sec:setup} below) and only the physical branch is shown for the linearized relation. 

In the Boltzmann equations we will solve, the difference in the equation of state between the 
two phases is actually not a bag constant but a mean field term and a bag constant, so 
\be
p_+ = \frac{a}{3} T^4 -\epsilon, \quad p_- = \frac{a}{3} T^4 - b \, m^2 T^2 \, ,
\ee
with $a$, $b$ and $\epsilon$ given by the SM particle content and Higgs potential. The phase transition will happen close to the critical temperature, so
\be
b m^2 T^2 \simeq \epsilon \, ,
\ee
since it is weak. This will be enforced by the Higgs equation in the Boltzmann setup. Defining the strength of the phase transition in terms of the trace of the energy-momentum tensor~\cite{Giese:2020rtr},
\be
\alpha = \frac{\Delta \theta}{3 w_+}
= \frac{2 \epsilon  - b \, m^2 T_-^2}{2 a\, T_+^4} \simeq \frac{b \, m^2}{2a\, T_+^2},
\ee
then leads again to the result (\ref{eq:vpvmsing}) that was obtained with only a bag constant.
We will see later that one can recover this relationship in the linearized Boltzmann equations. 

In summary, the physical solution shows a discontinuity close to the speed of sound (or more precisely the Jouguet velocity) because the globally consistent solutions are changing the branch of solutions in (\ref{eq:vpvm}). Linearizing the 
 equation in the bag constant and the mean field term (which encapsulate the energy injection) then leads to a singularity in the solution for a wall velocity 
close to the speed of sound, which stems from the non-analyticity of the square root in (\ref{eq:cslimit}). 

The solution to the Boltzmann equation will be valid in the regime $\alpha \ll X_-^2$ and 
beyond this point the actual solution would saturate to the value in (\ref{eq:cslimit}) rather than to blow up.
One should 
also explicitly check that the solution to the Boltzmann equation selects the globally consistent branch,
meaning $v_+ \lessgtr v_-$  for $v_- \lessgtr c_s$. 
This is automatically fulfilled for the linearized system if it abides to (\ref{eq:vpvmsing}).

\section{Linearized Boltzmann setup}
\label{sec:setup}

The results from the hydrodynamic analysis discussed in section~\ref{sec:origin} can also be obtained from a microphysical perspective, studying the non-equilibrium dynamics of the plasma during the phase transition. For not-too-strong phase transitions, with $\alpha\ll 1$, the Higgs bubble wall is typically thicker than the de Broglie wavelength of plasma particles, and the non-equilibrium distribution function of particle species $i$ can be found by solving the semi-classical Boltzmann equation
\be
p^\mu \partial_\mu \, f_i(x^\mu, p^\mu) + 
m \, F^\mu \partial_{p^\mu} \, f_i(x^\mu, p^\mu)  + {\cal C}[f_j] = 0 \, ,
\label{eq:Boltzmann}
\ee
where $p^\mu$ is the four-momentum of the particles evaluated on-shell ($p^0 = E = \sqrt{\vec p^{\, 2} + m^2}$), $F^\mu$ is the semi-classical force from the wall driving the particles out of equilibrium, and ${\cal C}$ is the collision term which tends to bring the system back to equilibrium.

This is a partial integro-differential equation, and in order to make progress we need to make an \emph{Ansatz} for the $f_i$'s, such that we can actually solve the integrals in the collision term and reduce the problem to a set of ordinary differential equations for the fluctuations characterizing the displacement from equilibrium. We will postpone a detailed discussion on this issue until the next section, focusing here instead on the general argument which will allow us to recover the results of section~\ref{sec:origin}.

First, note that, in principle, there is one such set of equations for each particle species in the plasma. However, only the heavier species, such as the $W^\pm$'s, $Z^0$'s and top quarks, feel the passage of the bubble and have a significant non-vanishing source term. The other species, namely photons, gluons, light quarks and leptons, remain very close to mutual equilibrium among themselves, and characterize the so-called \emph{background}. They are collectively driven out of equilibrium due to collisions with the heavy particles, which communicate the passage of the bubble to the background. We will parametrize the fluctuations of the heavy species relative to the background fluctuations.

Following~\cite{Moore:1995si, Konstandin:2014zta} we can model the non-equilibrium functions $f_i$ as
\begin{equation}
	f_i(x,p) = \frac{1}{e^{\beta(p^\mu u_\mu - \delta_p - \delta_{p,\text{bg}})} \pm 1},
	\label{eq:flow}
\end{equation}
with $\beta^{-1}=T$ the temperature of the plasma and $u_\mu$ its four-velocity relative to the observer. Expanding the fluctuations in powers of momenta and keeping only first order terms one obtains
\be
    \delta_p^\text{perfect fluid} = 
    \delta\mu + p^\mu( \delta u_\mu -  u_\mu \delta T/T  ),
    \label{eq:defdelta}
\ee
and analogously for $\delta_{p,\text{bg}}$. The plasma four-vector  is in the wall frame given as $u^\mu = \gamma_w (1,v_w)$, while we 
parametrize $\delta u_\mu = \delta v \, \bar u_\mu = 
\delta v \, \gamma_w (v_w,1)$. 

Thus $\delta\mu$ encodes information on the chemical potential of the fluid composed of species $i$, $\delta T/T$ parametrizes its temperature deviation relative to the plasma, and $\delta v$ are fluctuations of the fluid velocity. Photon and gluon numbers quickly equilibrate which implies  $\delta\mu_\text{bg}=0$~\cite{Moore:1995si}, so the background is described only by temperature and velocity fluctuations.

To solve for the three fluctuations of gauge bosons and tops we take three moments of the Boltzmann equation, multiplying it by $1$, $p^\mu u_\mu$ and $p^\mu \bar u_\mu$ and integrating over momenta. The resulting equations correspond to conservation equations for particle number and energy-momentum~\cite{Konstandin:2014zta}. The \emph{Ansatz} allows for solving the collision terms and, after linearizing in the fluctuations, and assuming a planar wall (such that the problem becomes essentially one-dimensional in the coordinate $z$) we are left with a system, as seen in the rest frame of the wall, of the form~\cite{Konstandin:2014zta}
\be\begin{split}
 A_W \cdot (q_W + q_{bg})^\prime + \Gamma_W \cdot q_W &= S_W, \, \\
  A_t \cdot (q_t + q_{bg})^\prime + \Gamma_t \cdot q_t &=  S_t, \, \\
  A_{\text{bg}} \cdot q_{bg}^\prime + \Gamma_{\text{bg},W} \cdot q_W + \Gamma_{\text{bg},t} \cdot q_t &= 0,
\label{eq:smallsys}
\end{split}\ee
where $q = (\delta\mu, -\delta T/T, \delta v)^T$ and prime denotes derivative with respect to the coordinate $z$. The matrices $A_W$, $A_t$ and $A_\text{bg}$ take the form
\be
A_i= \frac{\gamma_w}{2\pi^2}\begin{pmatrix}
v_w c_2 & v_w c_3 & c_3/3 \\
v_w c_3 & v_w c_4 & c_4/3 \\
c_3/3 & c_4/3 & v_w c_4/3 \\
\end{pmatrix},\quad
A_\text{bg} = \gamma_w\frac{c_4^{\rm bg}}{2\pi^2} \begin{pmatrix}
    v_w & 1/3\\
    1/3 & v_w/3
\end{pmatrix},
\label{eq:kin_o1}
\ee
with coefficients $c_n$ differing for bosons and fermions (see section~\ref{sec:gen_fluid} below), and $\gamma_w = (1-v_w^2)^{-1/2}$ the relativistic factor associated to the wall velocity. The background matrix corresponds to the bottom right $2\times 2$ block of the $A_i$, summed over all fermionic and bosonic degrees of freedom in the background\footnote{The background contains 78 fermionic degrees of freedom ($5$ light quarks $\times\, 12$ d.o.f each, $3$ charged leptons $\times\, 4$ d.o.f and $3$ neutrinos $\times\, 2$ d.o.f each) and 19 bosonic (16 gluons, 2 photons and 1 Higgs).}, so  $c^{\rm bg}_4 = 78 c_4^f + 19 c_4^b$. The collision matrices
$\Gamma_W$, $\Gamma_t$ are then $3\times 3$ while $\Gamma_{bg,X}$ are $2\times3$. Notice that the matrices $A_i$ and $A_\text{bg}$ have a vanishing eigenvalue at $v_w=1/\sqrt{3}=c_s$.

Finally, the source terms are given by moments over the forces in the Boltzmann equation
\be
S = \gamma_w v_w \frac{ (m^{2})^\prime}{4\pi^2 T^2} \, 
\begin{pmatrix}
    c_1\\
    c_2\\
    0
\end{pmatrix} \, ,
\ee
and is present for all species that significantly change mass during the phase transition.

The singularity in the friction arises then from an interplay of energy-momentum conservation, the peculiar form of the collision terms, and the zero eigenvalue in the kinetic matrix. 
First, notice that the equation of motion of the two background degrees of freedom can be uniquely fixed using energy-momentum conservation. 
In other words, there are two vectors $\chi$ such that $\chi \cdot \Gamma = 0$ which relates the matrices $\Gamma_{\text{bg},X}$ to the corresponding $2\times3$ parts of the matrices 
$\Gamma_{X}$. The collision terms drop out in the corresponding equations
\be
\chi \cdot A \cdot \Delta q = \chi \cdot  \int S \, dz\, .
\label{eq:shifts}
\ee
These two equations represent the linearized form of the energy-momentum conservation (\ref{eq:Tconserv}). The source terms 
parametrize the difference in the equation of state between the two phases due to the mean field term, 
$p_+ - p_- \propto m^2 T^2 \propto \int S \, dz$. At the same time, almost all fluctuations $\Delta q$ are damped by the collision terms
and tend to zero away from the source. The sole exceptions are the two fluctuations that parametrize a collective shift in the local equilibrium,
parameterized by a temperature shift $\delta T$ and a shift in the fluid motion $\delta v$. 
Hence, away from the source, the constraint (\ref{eq:shifts}) becomes essentially a $2\times 2$ system involving only $\delta T_{\text{bg}}$ and $\delta v_{\text{bg}}$. This will produce a singularity in the fluctuations once the wall velocity approaches the speed of sound, in analogy to (\ref{eq:vpvmsing}).   

To be specific, the relation (\ref{eq:shifts}) can be solved for the change of the fluid velocity across the wall,
\be
\delta v_{\rm bg} =  \frac{\sum_i N_i m_i^2 \, c^i_2}{2T^2\sum \, c_4} 
\frac{3 v_w}{1 - 3 v_w^2} \, ,
\label{eq:shiftbg}
\ee
where the sum over $i$ runs over the heavy particles (tops, $W^\pm$ and $Z^0$) whereas the sum over $c_4$ runs over heavy particles as well as over background species. This exactly reproduces the result of the hydrodynamic relation (\ref{eq:vpvmsing}), due to 
\be
\alpha = \frac{b \, m^2}{2a \,  T^2} 
= \frac{\sum N_i m_i^2 \, c^i_2}{2T^2 \sum \, c_4} \, ,
\ee
which relates the thermodynamic potentials to the 
coefficients in the Boltzmann equations, and the relation $(v_- - v_+) \gamma^2 = \delta v_{\rm bg}$ that 
stems from $du^\mu/dv_w = \gamma^2 \bar u^\mu$ and the definition of $\delta v$ in (\ref{eq:defdelta}). For a fiducial phase transition strength $\phi_0/T=1$ this yields $\alpha\approx 0.002$.

Notice that this conclusion seems unavoidable as long as the linearized Boltzmann equations are solved, even if a generalized Ansatz is used and many more moments are taken. Still, the blow up of these perturbations point to a breakdown of the linearization procedure. This argument only relies on the facts that: a) two moments of the Boltzmann equation represent total energy-momentum conservation, and b) the Ansatz for the distribution functions contains fluctuations that represent collective changes in the local equilibrium (i.e.~for the collective temperature and fluid velocity).

\section{Generalized fluid \emph{Ansatz}}
\label{sec:gen_fluid}

Let us now consider a generalized fluid \emph{Ansatz}, where the distribution functions still have the form in~(\ref{eq:flow}) but now the fluctuations read
\be
\delta_p = w^{(0)} + p^\mu\,w^{(1)}_\mu + p^\mu p^\nu w^{(2)}_{\mu\nu}+\ldots
\label{eq:gen_fluid}
\ee
and analogously for $\delta_{p,\text{bg}}$. In other words, we expand the non-equilibrium fluctuations in powers of momenta. Keeping only terms up to linear order in momenta, as in section~\ref{sec:setup}, consists in a description of a perfect fluid. Our generalized \emph{Ansatz} allows for more general entropy-producing phenomena~\cite{DeGroot:1980dk, Moore:1995si}. 

Before proceeding with the analysis, a few comments are in order. First, note that~(\ref{eq:gen_fluid}) is not the most general Ansatz possible, and may not be appropriate in certain circumstances. In particular, additional tensor structures would be best suited for fully resolving the angular dependence of the 
distribution function in high energy interactions, such as
$p_i/|p_i|$, $p_i p_j/|p_i|^2$, and so on. Since we will always compute the collision terms at leading-log, implying small momentum exchange, the parametrization adopted in~(\ref{eq:gen_fluid}) should be adequate. We will confirm {\em a posteriori} that the friction converges using our set of fluctuations. Still, even if not the most general, using some \emph{Ansatz} for the $f_i$'s is absolutely indispensable, since otherwise the momentum dependence of the collision terms
cannot be resolved. Even in approaches that claim \emph{Ansatz}-independence, a choice has to be made for the collision terms, which is ultimately equivalent to choosing an \emph{Ansatz} for the 
distribution function. In such approaches one actually has to be careful that the implicit \emph{Ansatz} chosen for the collision terms is consistently applied in the other terms of the Boltzmann equation as well.  For instance, in~\cite{Cline:2020jre} the collision terms were computed using the perfect fluid \emph{Ansatz} discussed in section~\ref{sec:setup}, whereas the kinetic term was computed differently. This consistency is difficult to accomplish without knowing the \emph{Ansatz} explicitly. Here it is guaranteed by construction.

\subsubsection*{Collision terms and background equations}

With the \emph{Ansatz}~(\ref{eq:gen_fluid}) the linearized collision term has the form
\be\begin{split}
    \mathcal{C}[f] =&\, \frac{1}{2T}\sum_{\rm processes} \int
				\dfrac{d^3k\, d^3p^\prime d^3k^\prime}{(2\pi)^{9} 2E_k\, 2E_{p^\prime}\, 2E_{k^\prime}} |\mathcal{M}_{pk\to p^\prime k^\prime}|^2 (2\pi)^4 \delta^4(p+k-p^\prime-k^\prime)\times\\
	& \times f_{0p}f_{0k}(1\pm f_{0p^\prime})(1\pm f_{0k^\prime})\, \bigg[(\delta_p + \delta_k - \delta_{p^\prime} - \delta_{k^\prime}) + 
	(\delta_{p,\text{bg}} + \delta_{k,\text{bg}} - \delta_{p^\prime,\text{bg}} - \delta_{k^\prime,\text{bg}})\bigg].
	\label{eq:lin_coll}
\end{split}\ee

Now, an interesting cancellation takes place for some of the background perturbations, which is at the core of the singularity across the speed of sound for the background fluctuations. Recalling that we can set $\delta\mu_{\rm bg} = w^{(0)}_\text{bg}=0$ since photon and gluon chemical potentials equilibrate quickly~\cite{Moore:1995si}, we have
\be\begin{split}
    \delta_{p,\text{bg}} + \delta_{k,\text{bg}} - \delta_{p^\prime,\text{bg}} - \delta_{k^\prime,\text{bg}} &=
    w^{(1)}_{\mu,\text{bg}}\,(p^\mu + k^\mu - p^{\prime\,\mu}-k^{\prime\,\mu}) \\
    &\quad + w^{(2)}_{\mu\nu,\text{bg}} (p^\mu p^\nu + k^\mu k^\nu - p^{\prime\,\mu}p^{\prime\,\nu} - k^{\prime\,\mu}k^{\prime\,\nu}) + \ldots,
    \label{eq:damped}
\end{split}\ee
and we see that the term proportional to $w^{(1)}_\mu$ vanishes due to energy-momentum conservation. This means that the leading-order fluctuations in the background equations are undamped. This cancellation does not happen for higher order fluctuations, which involve more complicated momentum dependences. In fact, we expect them to be severely damped, because of the large number of processes involving background species. In other words, these fluctuations equilibrate quickly, and we can set them to zero from the start. Thus, even if we adopt an extended \emph{Ansatz} for the heavy species, the background can be modelled by two perturbations only, corresponding to the undamped $w^{(1)}_\mu$ --- the background behaves as a perfect fluid.

\begin{figure}
    \centering
    \includegraphics[page=1]{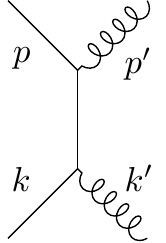}\quad
    \includegraphics[page=2]{figs/diagrams.pdf}\quad
    \includegraphics[page=3]{figs/diagrams.pdf}\quad
    \includegraphics[page=4]{figs/diagrams.pdf}\\[2mm]
    \includegraphics[page=5]{figs/diagrams.pdf}\quad
    \includegraphics[page=6]{figs/diagrams.pdf}\quad
    \includegraphics[page=7]{figs/diagrams.pdf}\quad
    \includegraphics[page=8]{figs/diagrams.pdf}\quad
    \includegraphics[page=9]{figs/diagrams.pdf}
    \caption{Collision processes taken into account in this study. We compute only the leading-log contributions, which means focusing only on $t$ and $u-$channel diagrams. The four diagrams in the first row are processes changing the perturbations of the top quark, namely double annihillation into gluons, scattering by light quarks, scattering by gluons, and absorption and reemission of a gluon. The second row are processes changing the distribution of $W$'s, namely single annihillation (by gluons into quarks and by quarks into gluons), double annihilation into quarks, absorption and reemission, and scattering by quarks.}
    \label{fig:diagrams}
\end{figure}

We will consider the collision processes shown in figure~\ref{fig:diagrams}, involving interactions of the heavy species with background particles. In these diagrams the heavy species are incoming particles in the interaction, so the corresponding collision term will enter the equation for this species. Taking moments of the Boltzmann equation will introduce a factor $p^\mu p^\nu\ldots$, with $p$ the momentum of this species. Importantly, for the two equations corresponding to energy-momentum conservation, there will be just a single power of $p^\mu$. If we now consider the equation for the other species in the diagram, this external momentum becomes $p\rightarrow k, p^\prime$ or $k^\prime$, and we then get the corresponding collision term appearing in the background equations. So when summing over the equations for all species, a factor $p+k-p^\prime-k^\prime$ arises in the energy-momentum equations, and the result vanishes. This argument does not rely on the specific Ansatz for the distribution functions, and in general the moments of the Boltzmann equation that represent energy-momentum conservation do not contain collision terms. So, overall, we do not need to compute additional collision terms other than $\Gamma_W$ and $\Gamma_t$ appearing in the equations for the heavy species, because the corresponding matrices for the background equations will be 
\be
    \Gamma_{\text{bg},W} = -N_W \Gamma_W,
    \qquad
    \Gamma_{\text{bg},t} = -N_t\Gamma_t.\, ,
    \label{eq:Gbg}
\ee
for the two moments that correspond to energy-momentum conservation. Since we deal with only two fluctuations for the background, these are all the collision terms we need to solve the system\footnote{This is the justification of the statement in section~\ref{sec:setup}, that there are two vectors $\chi$ corresponding to energy-momentum conservation such that $\chi\cdot \Gamma=0$ for the overall collision matrix $\Gamma$. }.

Truncating the expansion~(\ref{eq:gen_fluid}) to keep terms up to order $\ell$ in powers of momenta (the usual perfect fluid \emph{Ansatz} corresponding to $\ell=1$) results in a description with  $N=(\ell+1)(\ell+2)/2$ fluctuations. We then need to take $N$ independent moments of the Boltzmann equation, which is obtained by multiplying~(\ref{eq:Boltzmann}) by factors $(p^\mu u_\mu)^m (p^\mu \bar{u}_\mu)^n$, with $0\leq m+n\leq N$, and integrating over momenta. The resulting matrices appearing in the system~(\ref{eq:smallsys}) to an arbitrary order $\ell$ are obtained from integrals given in the appendix of reference~\cite{Dorsch:2021ubz}. At $\ell=2$, with $N=6$ fluctuations arranged in a vector as $q = (w^{(0)}, w^{(1)}_0, w^{(1)}_z, w^{(2)}_{00}, w^{(2)}_{0z}, w^{(2)}_{zz})^T$, the Boltzmann equation becomes a matrix system as in~(\ref{eq:smallsys}), with kinetic terms
\be
	A_i = \frac{\gamma_w}{2\pi^2}\begin{pmatrix}
		v_w c_2 & v_w c_3 & c_3/3 & v_w c_4 & c_4/3 & v_w c_4/3\\
		v_w c_3 & v_w c_4 & c_4/3 & v_w c_5 & c_5/3 & v_w c_5/3\\
		c_3/3 & c_4/3 & v_w c_4/3 & c_5/3 & v_wc_5/3 & c_5/5\\
		v_w c_4 & v_w c_5 & c_5/3 & v_w c_6 & c_6/3 & v_w c_6/3 \\
		c_4/3 & c_5/3 & v_w c_5/3 & c_6/3 & v_w c_6/3 & c_6/5 \\
		v_w c_4/3 & v_w c_5/3 & c_5/5 & v_w c_6/3 & c_6/5 & v_w c_6/5
	\end{pmatrix},
	\label{eq:kingen}
\ee
where, for $n\geq 2$,
\bea
c_n^b \equiv \frac{1}{T^{n+1}}\int dp \, p^n  f^\text{BE}_p (1+ f^\text{BE}_p) &=& n!\,\zeta_n\,,\\
c_n^f \equiv \frac{1}{T^{n+1}}\int dp \, p^n  f^\text{FD}_p (1- f^\text{FD}_p) &=& \left(1-\frac{1}{2^{n-1}}\right) n!\,\zeta_n\,,
\label{eq:p_ints}
\eea
and $f^{\text{BE},\, \text{FD}}$  the Bose-Einstein and Fermi-Dirac distributions. For $n=1$ one has $c_1^f = \log2$ and $c_1^b=\log(2T/m)$.

With gauge couplings $\alpha_s \approx 0.12$ and $\alpha_W \approx 1/30$, the collision matrices are
\be
    \Gamma_t = \begin{pmatrix}
        0.00899 &  0.01752 &  0 & 0.05489 & 0 & 0.01830 \\
        0.01752 &  0.05311 &  0 & 0.234580 &  0 &  0.07819 \\
        0 & 0 & 0.01801 &  0 &
        0.08331 \\
       0.05489 & 0.23458 & 0 & 1.38413 & 0 & 0.46138 \\
       0 & 0 & 0.08331 & 0 & 0.51188 & 0 \\
       0.01830 & 0.07819 & 0 & 0.46138 & 0 & 0.34591
    \end{pmatrix}T,
\ee
\be
    \Gamma_W = \begin{pmatrix}
        0.00466 & 0.00903 & 0 & 0.02822 & 0 & 0.00941\\
       0.00903 & 0.03101 & 0 & 0.13043 & 0 & 0.04348\\
       0 & 0 & 0.0123 & 0 & 0.0479 &  0\\
       0.02822 & 0.13043 & 0 & 0.71107 & 0 & 0.23702\\
       0 & 0 & 0.0479 & 0 & 0.25481 & 0\\
       0.00941 & 0.04348 & 0 & 0.23702 & 0 & 0.15841
    \end{pmatrix}T.
\ee
These numbers correct some mistakes in reference~\cite{Moore:1995si} that are mentioned in~\cite{Arnold:2000dr}.

Finally, the source term for fluctuations of a heavy species reads
\be
	{S} = \gamma_w v_w\frac{(m^2)^\prime}{4\pi T^2}
		\begin{pmatrix}
			c_1 & c_2 & 0 & c_3 & 0 & c_3/3		
		\end{pmatrix}^T.
\ee

With this setup, we solve the Boltzmann equation for the fluctuations using the shape \emph{Ansatz}
\be
    \phi(z) = \frac{\phi_0}{2}\left(1-\tanh \frac{z}{L_w}\right),
\ee
which gives the functional dependence $m(z)$ and its derivative. We can then plug the fluctuations into the Higgs equation to find the friction.

\section{Friction}
\label{sec:friction}

The passage of the bubble drives the plasma away from equilibrium, which then backreacts on the Higgs field, acting as a friction against the bubble expansion. Indeed, from energy-momentum conservation one can show that the Higgs equation of motion takes the form~\cite{Konstandin:2014zta}
\be
    -\phi^{\prime\prime} + \frac{dV_T}{d\phi} + \sum_i \frac{d m_i^2}{d\phi}\int \frac{d^3 p}{(2\pi)^3 2E_i} \delta f_i(x^\mu, p^\mu)=0,
    \label{eq:KG}
\ee
with $V_T$ the Higgs thermal potential in equilibrium. We see that the non-equilibrium contribution from $\delta f_i$ corresponds to a dissipative term, a friction which counters the bubble expansion.

There are two independent variables to be solved for, namely $v_w$ and $L_w$, so we take two moments of~(\ref{eq:KG}), multiplying it by $\phi^\prime$ and by $(2\phi-\phi_0)\times \phi^\prime$ and integrating. This yields
\be
    \frac{\Delta V_T}{T^4} = f_\text{fl} + f_\text{bg},
    \label{eq:p}
\ee
\be
-\frac{2}{15 (L_w T)^2}\left(\frac{\phi_0}{T}\right)^3 + \frac{1}{T^5}\int_0^{\phi_0} \frac{dV_T}{d\phi} (2\phi-\phi_0) d\phi = g_\text{fl}+g_\text{bg},
\label{eq:gradp}
\ee
with $\Delta V_T = V_T(\phi_0,T)-V_T(0,T)$ the free energy released during the phase transition. Equation~(\ref{eq:p}) states the balance in total pressure on the wall, equating the inside pressure to that of friction against it. The second equation~(\ref{eq:gradp}) is a balance of pressure gradient behind and in front of the wall, which also has to vanish overall because otherwise the wall would stretch or compress, which does not happen in the stationary state.

\begin{figure}
    \centering
    \includegraphics[width=.45\textwidth]{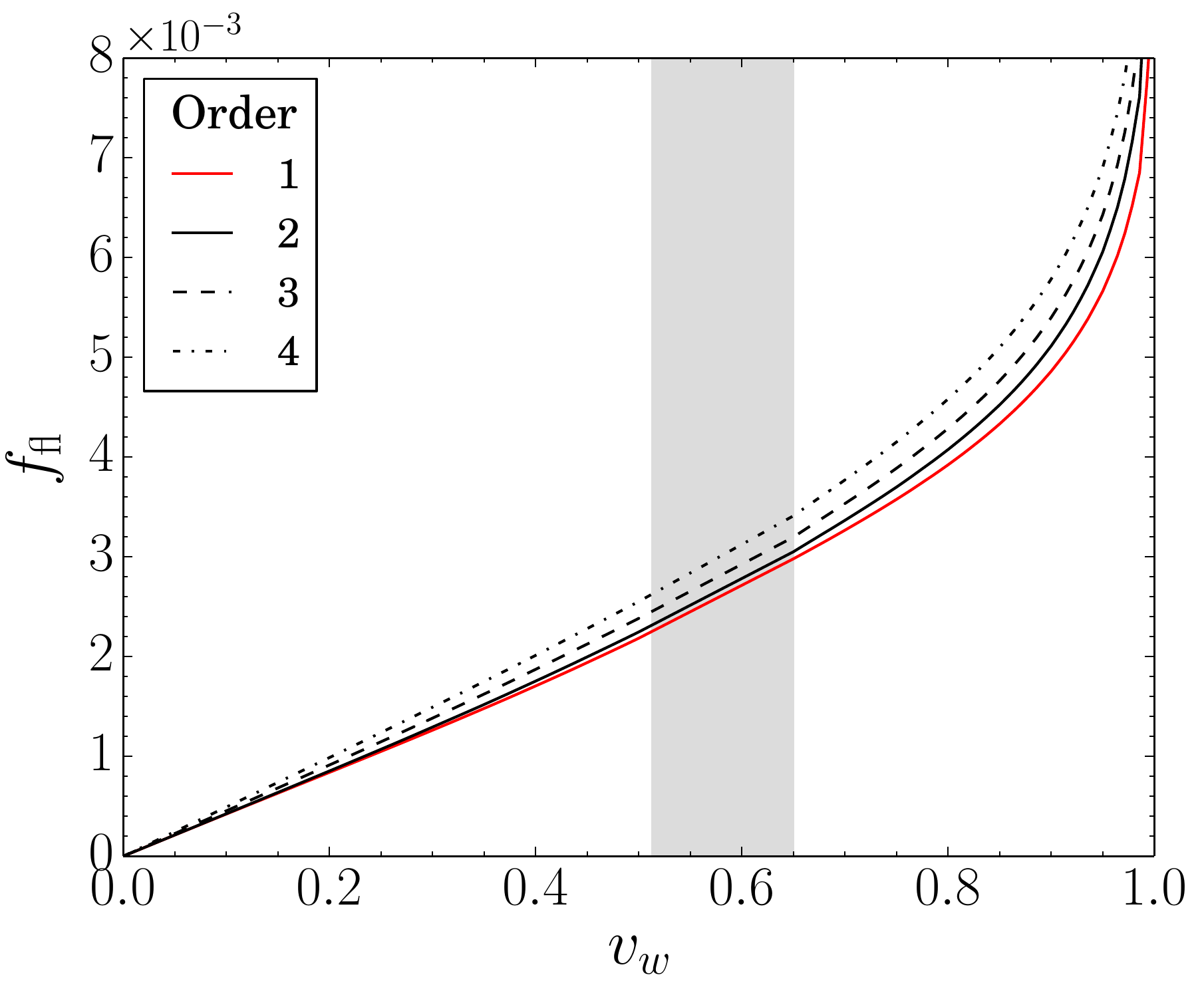}
    \includegraphics[width=.44\textwidth]{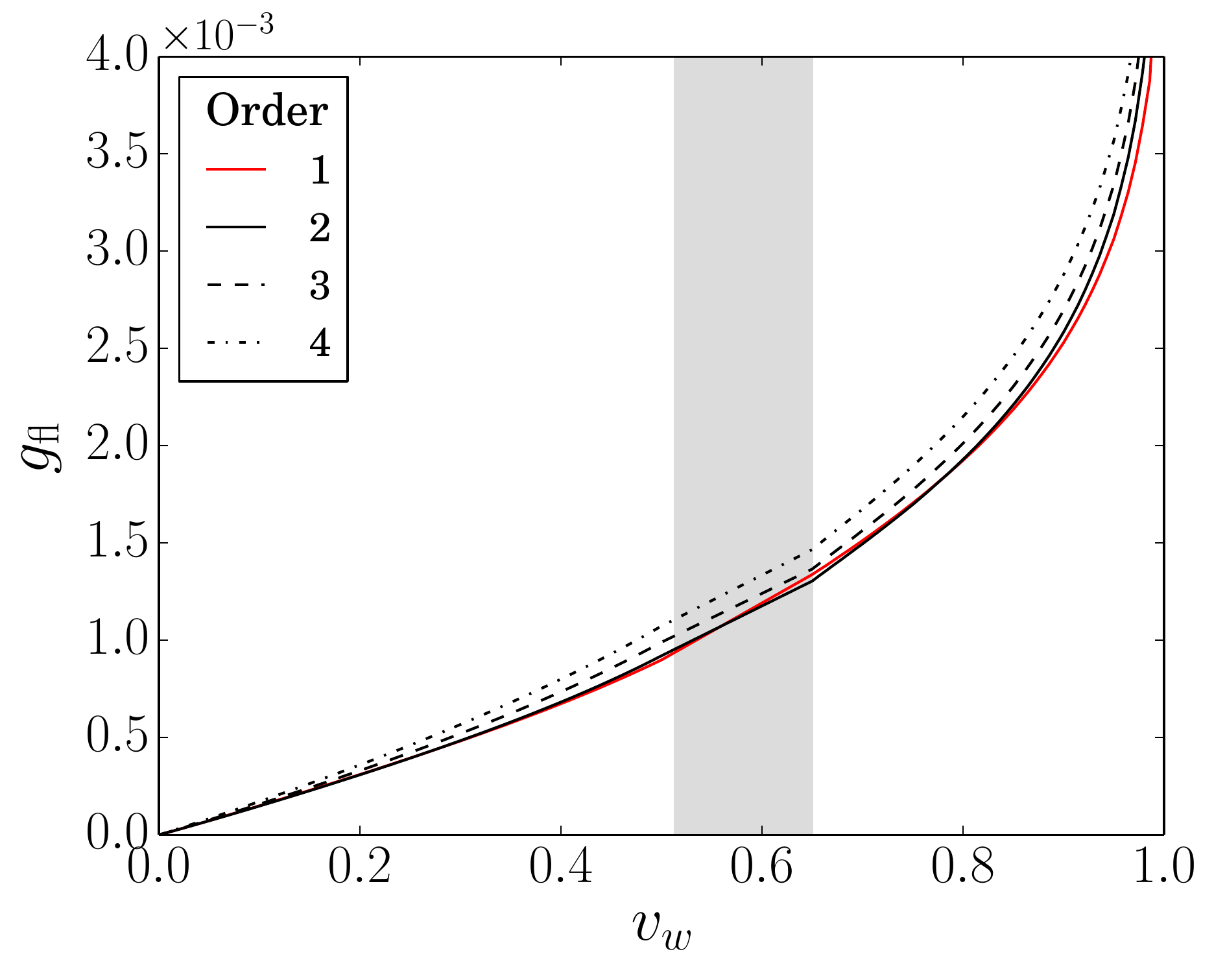}\\
    \includegraphics[width=.45\textwidth]{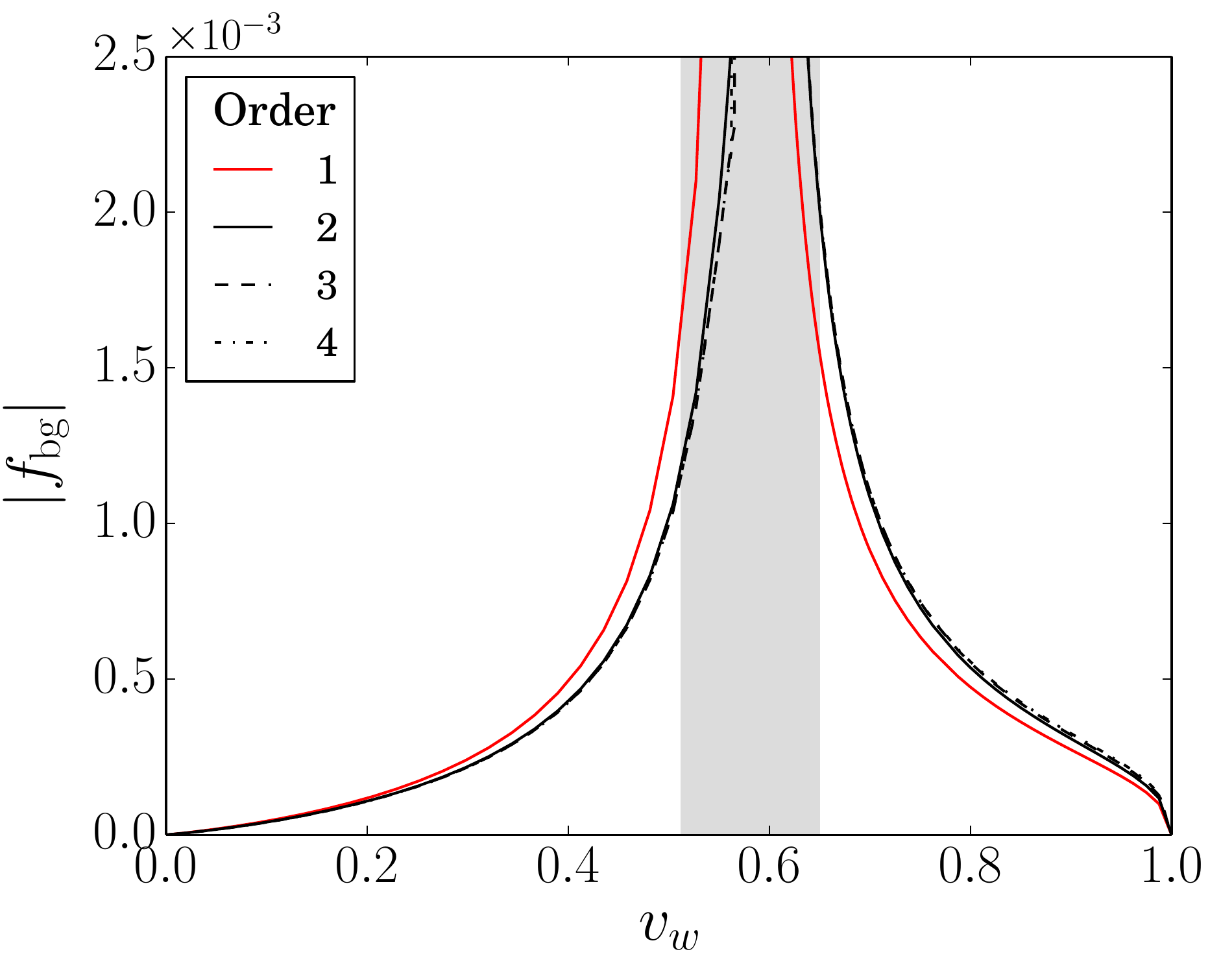}
    \includegraphics[width=.45\textwidth]{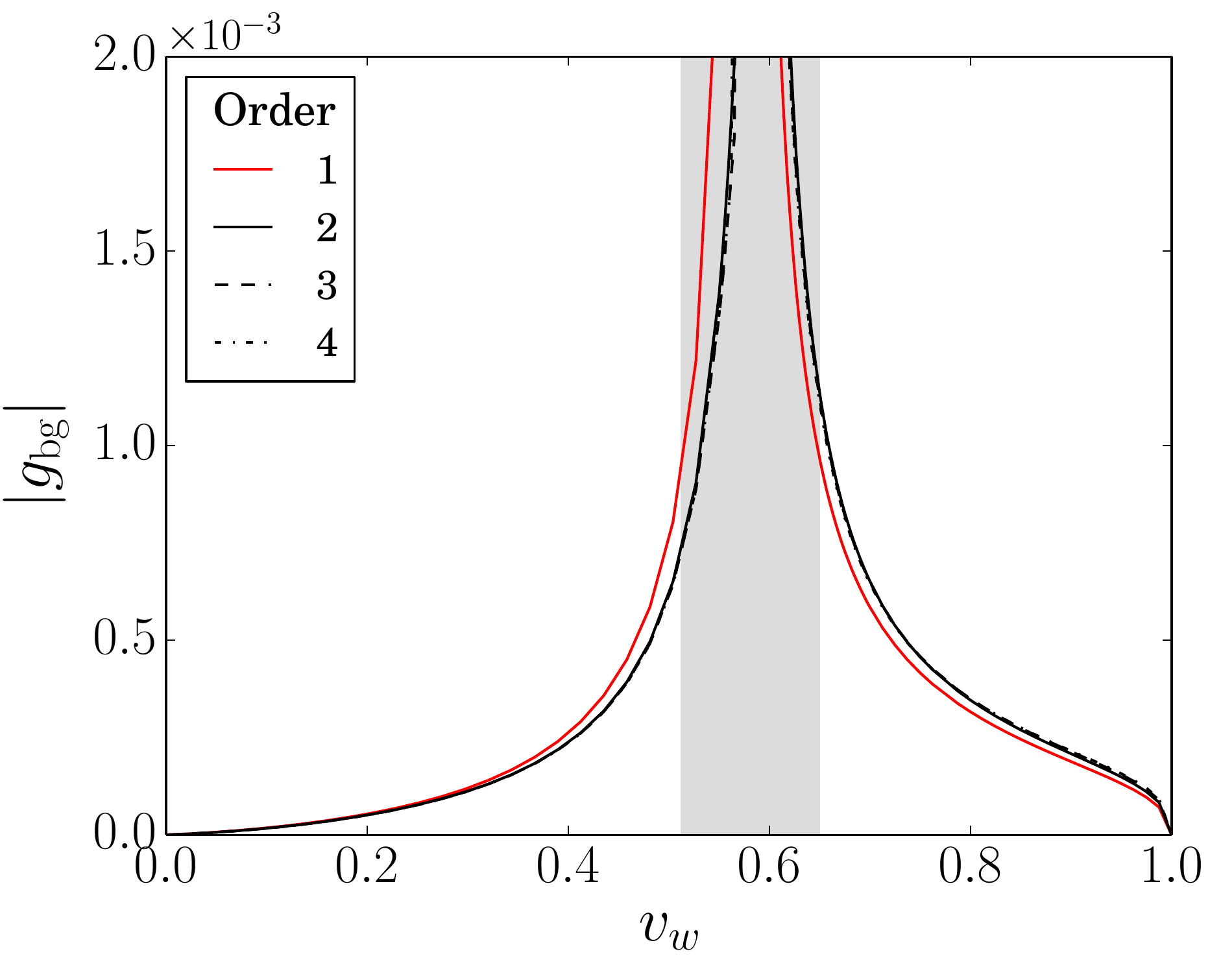}
    \caption{Friction contributions from the fluid and the background in equations~(\ref{eq:p}) and~(\ref{eq:gradp}) for different orders of the generalized fluid \emph{Ansatz}. Note that only the background contributions have a singularity, and that it appears only at the speed of sound $c_s=1/\sqrt{3}\approx 0.577$. We chose fiducial values for the wall thickness, $L_w T=10$, and the phase transition strength, $\phi_0/T = 1$, which yields $\alpha \approx 0.002$. The gray shaded regions satisfy $\alpha/X_-^2 > 1$, where we expect a breakdown of the linearization procedure, see section~\ref{sec:origin}.}
    \label{fig:fri}
\end{figure}

The functions $f_\text{fl}$ and $g_\text{fl}$ (respectively $f_\text{bg}$ and $g_\text{bg}$) are the contributions to friction coming from the fluctuations of the heavy species in the fluid (resp.~of the background). They are shown in figure~\ref{fig:fri} as a function of the wall velocity $v_w$, for various orders in momentum expansion of the fluid \emph{Ansatz} up to $\ell=5$, which takes into account a total of 21 fluctuations.

First, note that the friction contribution from fluctuations of the heavy species is non-singular at all velocities. This happens because, for these fluctuations, the Boltzmann system takes the general form
\be
    q^\prime + (A^{-1}\cdot\Gamma)\cdot q = A^{-1}\cdot S\,,
\ee
for some matrices $A$ and $\Gamma$ and a source $S$, whose solution behaves as
\be
    q(z)\sim \int^z A^{-1}\cdot S(z^\prime)\, e^{-\lambda z^\prime} dz^\prime.
    \label{eq:sol}
\ee
Here, $\lambda$ is an eigenvalue of $A^{-1}\cdot \Gamma$, which sets the rate at which  fluctuations decay. One sees that an eventual singular behaviour stemming from the kinetic matrix appears also in the exponential factor, effectively damping the singularity in $A^{-1}\cdot S$ and leading to the continuous behaviour seen in the plots. 

On the other hand, as previously discussed, the two background fluctuations $w^{(1)}_\text{bg}$ are not damped at all --- the collision terms vanish --- and one ends up with a solution similar to~(\ref{eq:sol}) but with $\lambda=0$, cf. equation~(\ref{eq:shifts}). Here, the vanishing eigenvalue in the kinetic matrix, which for these modes occur at $v_w=c_s$, leads to a singularity in the solutions and, consequently, in the friction functions $f_\text{bg}$ and $g_\text{bg}$.

Note that, as we increase the order in the fluid expansion, the kinetic matrix becomes singular for other values of the wall velocity. For instance, the matrix in~(\ref{eq:kingen}) is singular at $v_w=c_s$ and $v_w = \sqrt{3/5}$. However, these other singularities are associated with higher order fluctuations which \emph{are} damped by non-vanishing collision terms, cf. equations~(\ref{eq:lin_coll}) and~(\ref{eq:damped}). So we recover the continuous behaviour across these velocities. The only remaining singularity happens at $v_w=c_s$, which has a clear physical interpretation, as we argued in sections~\ref{sec:origin} and~\ref{sec:setup}.

We also remark that the friction functions do converge as we increase the number of fluctuations in the system, confirming that the series expansion in~(\ref{eq:gen_fluid}) is meaningful and well-behaved. In fact, beyond the second order the friction calculation seems very robust as long as the wall velocity is not too close to the speed of light. 

\section{Conclusions}
\label{sec:conclusions}

We revisited the friction calculation during first-order phase transitions in light of the recent developments in~\cite{Cline:2020jre} and~\cite{Laurent:2020gpg}. In the first part of the paper we argue that 
a discontinuity in the temperature and fluid background fields is expected on 
general grounds, just relying on hydrodynamic considerations. This stems from the fact that the global solutions to the fluid profiles transition from deflagrations to detonations when the wall velocity passes the Jouguet velocity.

We also show that this discontinuity turns into a singularity once the hydrodynamic equations are expanded in the (small) change in equation of state $\alpha$ that parametrizes the (weak) phase transition. This unphysical singularity comes from the fact that the hydrodynamic equations are non-analytic in $\alpha$ for wall velocities close to the speed of sound, $v_w \to c_s$, and $\alpha \to 0$, see (\ref{eq:cslimit}).

Based on this analysis, a singularity in the linearized Boltzmann equation is expected. Even if a method could be found to remedy the unphysical singularity for wall velocities close to the speed of sound, this method should still reproduce the discontinuity. 

At this point, it is worth comparing our results to the approach presented in~\cite{Laurent:2020gpg}, which finds a smooth behaviour for friction across the speed of sound. First, it should be noted that this continuity was actually used as an input for constructing the setup in~\cite{Laurent:2020gpg}, motivating the choices made for the moments and the fluctuations used in their \emph{Ansatz}. In particular, the moments of the Boltzmann equation taken in~\cite{Laurent:2020gpg} do not correspond to energy-momentum conservation. But this would imply that expression~(\ref{eq:Gbg}) for the background collision matrix,  which is assumed in~\cite{Laurent:2020gpg}, is no longer valid, and these collision terms would have to be computed separately.

Here we have shown that the interplay of energy-momentum conservation and the inclusion of fluctuations representing collective shifts in temperature and fluid velocity necessarily lead to a singularity close to the speed of sound. Energy-momentum conservation provides two conserved quantities while the collective shifts are not damped by collision terms. These features already imply the occurrence of a singularity, and we reproduce the exact expression for the singularity found in hydrodynamics starting from the Boltzmann equations, see (\ref{eq:vpvmsing}) and (\ref{eq:shiftbg}).  

Finally, we study the impact of a more general Ansatz for the fluctuations. In baryogenesis calculations, adding more fluctuations indeed changes the outcome qualitatively~\cite{Dorsch:2021ubz}, in agreement with~\cite{Cline:2020jre}. 
Here we find that generalizing the Ansatz only leads to minor changes in the final outcome for the friction, but the main qualitative feature --- the singularity across the speed of sound --- still remains.

\section*{Acknowledgements}

We thank Jim Cline and Benoit Laurent for useful discussions. GCD acknowledges support from Pr\'o-Reitoria de Pesquisa of Universidade Federal de Minas Gerais (UFMG) under grant number 28359. TK is supported by the Deutsche Forschungsgemeinschaft under Germany’s Excellence Strategy – EXC 2121 ``Quantum Universe“ – 390833306. 
SJH is supported by the Science and Technology
Research Council (STFC) under the Consolidated Grant
ST/T00102X/1.

\end{document}